\documentclass[aps,prb,twocolumn,showpacs]{revtex4}

\begin{document}

\preprint{p01jul99}

\title{Gossamer Superconductivity}

\author{R. B. Laughlin}

\affiliation{Department of Physics, Stanford University, Stanford,
             CA 94305}

\homepage[R. B. Laughlin: ]{http://large.stanford.edu}

\date{September 1, 2002}

\begin{abstract}
An new superconducting hamiltonian is introduced for which the exact
ground state is the Anderson resonating valence bond.  It differs from the
t-J and hubbard hamiltonians in possessing a powerful attractive
force. Its superconducting state is characterized by a full and intact
d-wave tunneling gap, quasiparticle photoemission intensities that are
strongly suppressed, a suppressed superfluid density, and an incipient
Mott-Hubbard gap.
\end{abstract}


\maketitle

It has been known since the early work of Uemura \cite{uemura} that the
superfluid density and transition temperature of underdoped cuprate
superconductors both vanish more-or-less linearly with doping and are
proportional.  The constant of proportionality is consistent with the
transition being an order-parameter phase instability analogous to the
Kosterlitz-Thouless transition in 2 dimensions \cite{emery}. This idea is
supported by numerous other experiments, including the optical sum rule
studies of Uchida \cite{uchida}, the giant proximity effect reported by
Decca {\it et al} \cite{drew}, and the recent heat-transport measurements
of Wang {\it et al} \cite{ong} showing superconducting vortex-like effects
above the transition temperature. The transport trends continue into the
insulator, where Ando \cite{ando} reports that high-temperature hall
effect consistent with a carrier density roughly proportional to doping.
At the same time, however, the d-wave gap in the quasiparticle spectrum
grows monotonically as the doping decreases and saturates at a value of
about 0.3 eV \cite{marshal, ding, rbl, timusk}.  This has led to
speculations that the tunneling pseudogap is the energy to make a
pre-formed Cooper pair, which then condenses into a superfluid at a lower
temperature. There is no direct evidence that the d-wave nodal structure
survives into the insulator, but there is circumstantial evidence for
this, notably the observation in La$_{2-x}$Sr$_x$CuO$_4$ by Yoshida {\it
et al} \cite{fujimori} of dispersing quasiparticle bands near the d-wave
node that become fainter as doping is reduced but do not shift or change
their velocity scale.  These bands are also detached from the lower
Hubbard band, and simply materialize in mid-gap as the doping is increased
from zero.

All of this behavior is consistent with the idea that the
superconductivity persists deep into the ``insulating'' state, coexists
with antiferromagnetism there, and fails to conduct only because its
long-range order is disrupted, presumably on account of its low superfluid
density \cite{dynes}.  There are many ways the latter could occur,
including impurity localization or crystallization of the order parameter,
for such a {\it gossamer superconductor} is physically equivalent to a
dilute gas of bosons and thus highly unstable.

The idea that the ``insulator'' might actually be a thin, ghostly
superconductor is implicit in the mathematics of the Anderson resonating
valence bond (RVB) \cite{rvb} worked out by various authors in the late
1980s \cite{gros, shiba, zhang} and further extended recently by
Paramehanti, Randeria, and Trivedi \cite{randeria}. Unfortunately, this
idea has always run afoul of a basic premise of RVB theory that
superconductivity should be a universal aspect of quantum
antiferromagnetism. This premise is both confusing and fundamentally
incorrect, as the conventional spin density wave ground state, which
contains no superconductivity, is a perfectly good prototype for a quantum
antiferromagnet.  The real issue is not whether all antiferromagnets are
superconductors but whether some of them are - {\it i.e.} whether there
exists a second kind of antiferromagnetism distinguished from the first by
a tiny background superfluid density. One would also like to know which
hamiltonians favor this second kind of state over the first.  It is not
just the hamiltonians that stabilize antiferromagnetism, for these simply
emphasize the aspects of the vacua that are the same and de-emphasize the
aspects that are different. It has been known since the early work of Hsu
\cite{hsu}, for example, that these two kinds of vacuum have almost
identical variational energies in the context of the t-J hamiltonian. This
is consistent with the recent numerical-variational studies of Sorella
{\it et al} \cite{sorella}, who report that the t-J model superconducts in
a region of its parameter space, even though previous numerical work on
the same model reported antiferromagnetic stripe ordering \cite{stripe}.  
Becca {\it et al} \cite{becca} have argued that the latter is an artifact
lattice anisotropy. However the important point it that sensitivity to
algorithmic detail and the inherent difficulty of determining the order,
reflected in lack of agreement among different groups, demonstrate that
models of this kind are highly conflicted and close to a quantum phase
transition \cite{conflict}.  In other words, by exaggerating the magnetism
these models confuse the issue rather than clarifying it. There is no
persuasive evidence for superconductivity in the hubbard model \cite{su,
hanke, imada, bulut}.

The purpose of this letter is to propose a new strategy for resolving the
cuprate dilemma.  Rather than struggle to diagonalize a conflicted
hamiltonian, we shall modify the equations of motion to stabilize the
gossamer superconductor.  The easiest way to do this is by postulating a
powerful attractive force between electrons, just as one would in any
other superconductor.  This solution is not unique, but it is
experimentally falsifiable through the excitation spectrum of the state,
which is model-dependent. Insofar as these properties match experiment,
which has yet to be seen, it would suggest that coulomb interactions are
not sufficient to explain cuprate superconductivity.

We consider a planar square lattice of sites $j$ on which electrons 
may sit.  The superconducting vacuum we wish to stabilize is
$| \Psi \! > = \Pi_\alpha \; | \Phi \! >$, where

\begin{equation}
| \Phi \! > = \prod_{\bf k}^N (u_{\bf k} + v_{\bf k} 
c_{{\bf k}\uparrow}^\dagger c_{-{\bf k}\downarrow}^\dagger ) | 0 \! >
\; \; \; ,
\end{equation}

\noindent
with $c_{{\bf k} \sigma} = N^{-1/2} \sum_j^N \exp( i {\bf k} \cdot
{\bf r}_j ) \; c_{j \sigma}$ as usual, and

\begin{equation}
\Pi_\alpha = \prod_j^N z_0^{(n_{j \uparrow} + n_{j \downarrow})/2}
\; ( 1 - \alpha_0 \; n_{j \uparrow} n_{j \downarrow} )
\; \; \; .
\end{equation}

\noindent
The Bardeen-Cooper-Schrieffer (BCS) pairing amplitudes satisfy

\begin{equation}
\left[ \begin{array}{cc}
\epsilon_{\bf k} - \mu & \Delta_{\bf k} \\
\Delta_{\bf k} & \mu - \epsilon_{\bf k}
\end{array} \right]
\left[ \begin{array}{c}
u_{\bf k} \\ v_{\bf k} \end{array} \right]
= E_{\bf k}
\left[ \begin{array}{c}
u_{\bf k} \\ v_{\bf k} \end{array} \right]
\end{equation}

\noindent
and are normalized by $u_{\bf k}^2 + v_{\bf k}^2 = 1$. They are related to
the hole doping $\delta$ by

\begin{equation}
\frac{1}{N} \sum_{\bf k} v_{\bf k}^2 = 1 - 
\frac{1}{N} \sum_{\bf k} u_{\bf k}^2 = \frac{1 - \delta}{2}
\; \; \; .
\end{equation}

\noindent
The important positive eigenvalue

\begin{equation}
E_{\bf k} = \sqrt{ ( \epsilon_{\bf k} - \mu )^2 + \Delta_{\bf k}^2 }
\label{dispersion}
\end{equation}

\noindent
is the energy to make a quasiparticle (either an electron or a hole)
when the parameter $\alpha_0$ is zero. The parameter $z_0$ is a fugacity
required to keep the electron density the same as one varies
$\alpha_0$.  Assuming the superconducting order parameter to be d-wave, so
there is no on-site pairing amplitude, the charge states of a site are
statistically independent and characterized by a fugacity $z$.  The
condition that the total charge on the site be $1 - \delta$, where
$\delta$ is the doping, is $[2z + 2 (1 - \alpha) z^2]/[1 + 2 z + (1 -
\alpha ) z^2] = 1 - \delta$, where $1 - \alpha = (a - \alpha_0)^2$, 
or

\begin{equation}
z = \frac{\sqrt{1 - \alpha ( 1 - \delta^2)} - \delta}
{(1 - \alpha) ( 1 + \delta)} = ( \frac{ 1 - \delta}{1 + \delta} )
\; z_0 \; \; \; .
\end{equation}

\noindent
The parameter $z_0$ is the factor by which z exceeds $(1 - \delta)/(1 +
\delta)$, its value for $\alpha_0 = 0$.

The hamiltonian is constructed using the BCS annihilation operators

\begin{equation}
b_{{\bf k} \uparrow} = u_{\bf k} c_{{\bf k} \uparrow} +
v_{\bf k} c_{- {\bf k} \downarrow}^\dagger
\; \; \; \; \; \; \; \;
b_{{\bf k} \downarrow} = u_{\bf k} c_{{\bf k} \downarrow} -
v_{\bf k} c_{- {\bf k} \uparrow}^\dagger
\; \; \; .
\end{equation}

\noindent
So long as $\alpha_0 \neq 1$, the partial projector $\Pi_\alpha$ has an
inverse

\begin{equation}
\Pi_\alpha^{-1}
= \prod_j^N z_0^{- (n_{j \uparrow} + n_{j \downarrow})/2}
\; ( 1 + \beta_0 \; n_{j \uparrow} n_{j \downarrow} )
\; \; \; ,
\end{equation}

\noindent
where $\beta_0 = \alpha_0/(1 - \alpha_0)$.  This enables us to construct
the modified annihilation operators

\begin{displaymath}
\tilde{b}_{{\bf k} \uparrow} = \Pi_\alpha b_{{\bf k} \uparrow}
\Pi_\alpha^{-1}
= \frac{1}{\sqrt{N}} \sum_j^N e^{i {\bf k} \cdot {\bf r}_j}
\end{displaymath}

\begin{equation}
\times
\biggl[ z_0^{-1/2} u_{\bf k} 
(1 + \beta_0 n_{j \downarrow} ) c_{j \uparrow}
+ z_0^{1/2} v_{\bf k} 
(1 - \alpha_0 n_{j \uparrow} ) c_{j \downarrow}^\dagger
\biggr]
\; \; \; ,
\end{equation}

\noindent
and likewise for $\tilde{b}_{{\bf k} \downarrow}$, for which
$\tilde{b}_{{\bf k} \sigma} | \Psi \! > = 0$. Thus $| \Psi \! >$
is an eigenstate of the hamiltonian

\begin{equation}
{\cal H} = \sum_{{\bf k} \sigma} E_{\bf k}
\tilde{b}_{{\bf k} \sigma}^\dagger
\tilde{b}_{{\bf k} \sigma}
\label{hamiltonian}
\end{equation}

\noindent
with eigenvalue 0.  However, this hamiltonian has only
non-negative eigenvalues, since for any wavefunction $| \chi \!>$

\begin{equation}
< \! \chi | {\cal H} | \chi \! > = \sum_{{\bf k} \sigma} E_{\bf k}
< \! \tilde{b}_{{\bf k} \sigma} \chi |
\tilde{b}_{{\bf k} \sigma} \chi \! > \; \geq \; 0
\; \; \; .
\end{equation}

\noindent
Thus $| \Psi \! >$ is a ground state of ${\cal H}$.  However, it is
also {\it the} ground state by virtue of adiabatic continuity.  The
state in question may be continuously deformed into a BCS state by
taking $\alpha_0$ slowly to zero.  Since it does not cross a phase
boundary in the process, the ground state and low-lying excitations must
track in a one-to-one way.

Let us now consider the quasiparticle excitations of this superconductor.
The operators $\tilde{b}_{{\bf k} \sigma}$ no longer anticommute properly
with their hermitian adjoints and thus cannot be used to create
quasiparticles. The physical meaning of this is that the quasiparticles
interact.  Instead we will use the variational wavefunctions
$| {\bf k} \sigma \! > = \Pi_\alpha b_{{\bf k} \sigma}^\dagger
| \Phi \! >$ borrowed from the RVB literature. The expected energy is

\begin{equation}
\frac{< \! {\bf k} \sigma | {\cal H} | {\bf k} \sigma \! >}
{ < \! {\bf k} \sigma | {\bf k} \sigma \! >}
= E_{\bf k} \; \frac{ < \! \Psi | \Psi \! >}
{< \! {\bf k} \sigma | {\bf k} \sigma  \! >}
\simeq E_k \; \; \; .
\end{equation}

\noindent
The last step requires evaluating the relevant norms, which can be 
done by hand only approximately \cite{hsu, randeria}.  We repeat
the arguments here for completeness.  From

\begin{equation}
b_{{\bf k} \uparrow}^\dagger | \Phi \! >
= \frac{1}{u_{\bf k}^2} c_{ {\bf k} \uparrow}^\dagger | \Phi \! >
= \frac{1}{v_{\bf k}^2} c_{- {\bf k} \downarrow} | \Phi \! >
\end{equation}

\noindent
we find that

\begin{displaymath}
\frac{1}{N} \sum_{\bf k}^N u_k^2
\frac{< \! \Phi | b_{{\bf k} \uparrow} \Pi_\alpha^2
b_{{\bf k} \uparrow}^\dagger | \Phi \! >}
{ < \! \Phi | \Pi_\alpha^2 | \Phi \! >}
= \frac{< \! \Phi | c_{j \uparrow} \Pi_\alpha^2
c_{j \uparrow}^\dagger | \Phi \! >}
{< \! \Phi | \Pi_\alpha^2 | \Phi \! >}
\end{displaymath}

\begin{equation}
= z_0 \biggl[ \frac{1 + (1 - \alpha) z}{1 + 2 z + (1 - \alpha ) z^2}
\biggr] = \frac{1 + \delta}{2}
\end{equation}

\noindent
and

\begin{displaymath}
\frac{1}{N} \sum_{\bf k}^N v_k^2
\frac{< \! \Phi | b_{{\bf k} \uparrow} \Pi_\alpha^2
b_{{\bf k} \uparrow}^\dagger | \Phi \! >}
{ < \! \Phi | \Pi_\alpha^2 | \Phi \! >}
= \frac{< \! \Phi | c_{j \uparrow}^\dagger \Pi_\alpha^2
c_{j \uparrow} | \Phi \! >}{< \! \Phi | \Pi_\alpha^2 | \Phi \! >}
\end{displaymath}

\begin{equation}
= \frac{1}{z_0} \biggl[ \frac{z + z^2}{1 + 2 z + (1 - \alpha ) z^2}
\biggr] = \frac{1 - \delta}{2}
\; \; \; .
\end{equation}

\noindent
For $j \neq j'$ we assume that the amplitude for a given
configuration is weighted by the square root of its corresponding
probability, and that these weights add equally.  We then have

\begin{displaymath}
\frac{< \! \Phi | c_{j \uparrow} \Pi_\alpha^2
c_{j' \uparrow}^\dagger | \Phi \! >}
{< \! \Phi | \Pi_\alpha^2 | \Phi \! > } \times
\frac{ < \! \Phi | \Phi \! >}
{< \! \Phi | c_{j \uparrow} c_{j' \uparrow}^\dagger | \Phi \! >}
\end{displaymath}

\begin{equation}
\simeq \frac{4}{1 - \delta^2} (z z_0) \biggl[
\frac{1 + (1 - \alpha) z}{1 + 2 z + (1 - \alpha ) z^2} \biggr]^2 = 1
\end{equation}

\noindent
and

\begin{displaymath}
\frac{< \! \Phi | c_{j \uparrow}^\dagger \Pi_\alpha^2
c_{j' \uparrow} | \Phi \! >}
{< \! \Phi | \Pi_\alpha^2 | \Phi \! > } \times
\frac{ < \! \Phi | \Phi \! >}
{< \! \Phi | c_{j \uparrow}^\dagger c_{j' \uparrow} | \Phi \! >}
\end{displaymath}

\begin{equation}
\simeq \frac{4}{1 - \delta^2} (\frac{z}{z_0}) \biggl[
\frac{1 + z}{1 + 2 z + (1 - \alpha ) z^2} \biggr]^2 = 1
\; \; \; .
\end{equation}

Let us now consider the low-energy spectroscopic properties of this model.  
Reasoning as above, we find the matrix element for photoemission to be

\begin{equation}
\frac{ < \! - {\bf k} \downarrow | c_{{\bf k} \uparrow} |
\Psi \! > }
{\sqrt{< \! - {\bf k} \downarrow | - {\bf k} \downarrow > \;
< \! \Psi | \Psi \! > }} = g v_{\bf k}
\; \; \; ,
\end{equation}

\noindent
where

\begin{equation}
g^2 \simeq \frac{2 \alpha_0}{\alpha} \biggl\{ 1 - \frac{\alpha_0}{\alpha}
\biggl[ \frac{1 - \sqrt{1 - \alpha ( 1 - \delta^2)}}
{1 - \delta^2} \biggr] \biggr\}
\; \; \; .
\end{equation}

\noindent
Inverse photoemission is the same except with $u_{\bf k}$ substituted for
$v_{\bf k}$. The suppression of the photoemission intensity is matched by
a similar suppression of the superfluid order parameter:

\begin{equation}
\frac{< \! \Psi | c_{j \uparrow } c_{j' \downarrow} | \Psi \! >}
{< \! \Psi | \Psi \! >} \simeq g^2
\frac{< \! \Phi | c_{j \uparrow } c_{j' \downarrow} | \Phi \! >}
{< \! \Phi | \Phi \! >}
\; \; \; .
\end{equation}

\noindent
We need not consider the case of $j = j'$ for a d-wave superconductor.
Thus under strong projection near half-filling this model exhibits the
pseudogap phenomenon: The quasiparticle energies remain at their
unperturbed values $E_{\bf k}$ as $\alpha_0$ increases from 0 to 1, but
the superfluid density decreases from 1 to $2|\delta|/(1 + |\delta|)$.

Let us now consider the formation of the Mott-Hubbard gap.  Photoemission
of a quasiparticle accounts for only a small fraction of the sum rule

\begin{equation}
\frac{< \! \Psi | c_{{\bf k} \sigma}^\dagger c_{{\bf k} \sigma}
| \Psi \! >}{< \! \Psi | \Psi \! >} \simeq g^2 v_{\bf k}^2 +
(1 - g^2) \frac{1 - \delta}{2}
\; \; \; .
\end{equation}

\noindent
The rest must occur at a higher energy scale, the value of which may
be estimated by computing the expected energy of a hole.  From
the anticommutators

\begin{equation}
\{ \tilde{b}_{{\bf k} \uparrow} , c_{j \uparrow} \}
= \frac{1}{\sqrt{N}} e^{i {\bf k} \cdot {\bf r}_j} \biggl[
- z_0^{1/2} \alpha_0 v_{\bf k}
c_{j \uparrow} c_{j \downarrow}^\dagger \biggr]
\end{equation}

\begin{displaymath}
\{ \tilde{b}_{{\bf k} \downarrow} , c_{j \uparrow} \}
= \frac{1}{\sqrt{N}} e^{i {\bf k} \cdot {\bf r}_j} \biggl[
- z_0^{-1/2} \beta_0 u_{\bf k}
c_{j \uparrow} c_{j \downarrow}
\end{displaymath}

\begin{equation}
+ z_0^{1/2} v_{\bf k} ( 1 - \alpha_0 n_{j \downarrow} ) \biggr]
\end{equation}

\noindent
we obtain

\begin{displaymath}
< \! \Psi | c_{{\bf k} \sigma}^\dagger {\cal H} c_{{\bf k} \sigma} |
\Psi \! > = z_0 \biggl[ 1 - \alpha_0 \frac{1 - \delta}{2} \biggr]^2
v_{\bf k}^2 E_{\bf k}
\end{displaymath}

\begin{equation}
+ \frac{1}{N} \sum_{\bf q}^N E_{{\bf k} + {\bf q}} \biggl[
z_0 \alpha_0^2 A_{\bf q} v_{{\bf k} + {\bf q}}^2 +
\frac{\beta_0^2}{\alpha_0} B_{\bf q} u_{{\bf k} + {\bf q}}^2 \biggr]
\; \; \; ,
\end{equation}

\noindent
where 

\begin{displaymath}
A_{\bf q} = \sum_j^N \biggl[
\frac{< \! \Psi | {\bf S}_0 \cdot {\bf S}_j | \Psi \! >}
{< \! \Psi | \Psi \! >}
\end{displaymath}

\begin{equation}
+ \frac{1}{4} \frac{< \! \Psi | n_0 n_j | \Psi \! >}
{< \! \Psi | \Psi \! >} - \frac{(1 - \delta)^2}{4}
\biggr] e^{i {\bf q} \cdot {\bf r}_j}
\end{equation}

\begin{equation}
B_{\bf q} = \sum_j^N
\frac{< \! \Psi | c_{0 \uparrow}^\dagger c_{0 \downarrow}^\dagger
c_{j \downarrow} c_{j \uparrow} | \Psi \! >}
{< \! \Psi | \Psi \! >} e^{i {\bf q} \cdot {\bf r}_j}
\; \; \; .
\end{equation}

\noindent
Proceeding similarly with the state $c_{{\bf k} \uparrow}^\dagger | \Psi
\! >$, we find that the energy to inject an electron is the exact
particle-hole conjugate of this expression, produced from it by
interchanging $u_{\bf k}$ and $v_{\bf k}$, negating $\delta$, and
substituting of $1 - n_j$ for $n_j$.  Thus at half-filling the electron
spectral function is symmetric. The density and superfluid correlations
become suppressed at half-filling in the $\alpha_0 \rightarrow 1$ limit,
while the magnetic correlations become enhanced.   Approximating the
latter by the correlation function of the N\'{e}el vacuum, we obtain at
$\delta = 0$

\begin{displaymath}
\lim_{\alpha_0 \rightarrow 1}
\frac{< \! \Psi | c_{{\bf k} \sigma}^\dagger {\cal H} c_{{\bf k} \sigma}
| \Psi \! >}
{< \! \Psi | c_{{\bf k} \sigma}^\dagger c_{{\bf k} \sigma} | \Psi \! >}
= \lim_{\alpha_0 \rightarrow 1}
\frac{< \! \Psi | c_{{\bf k} \sigma} {\cal H} c_{{\bf k} \sigma}^\dagger
| \Psi \! >}
{< \! \Psi | c_{{\bf k} \sigma} c_{{\bf k} \sigma}^\dagger \Psi \! >}
\end{displaymath}

\begin{equation}
\simeq
\frac{1}{1 - \alpha_0} \biggl[ \frac{1}{N} \sum_{\bf q}^N E_{\bf q}
+ \frac{1}{2} E_{\bf k} \biggr]
\; \; \; .
\end{equation}

\noindent
Thus the spectral function consists of Mott-Hubbard ``lobes'' at high
energies with a faint band of states at mid-gap associated with the
gossamer quasiparticles.  The chemical potential does not jump in this
model when $\delta$ is tuned from negative to positive, as it does in the
hubbard model \cite{bulut}.

Let us now consider antiferromagnetism. At half-filling the largest term
in Eq. (\ref{hamiltonian}) is an on-site coulomb repulsion of magnitude $2
\sum_{\bf k}^N E_{\bf k}/N (1 - \alpha_0)$. If this repulsion is made
slightly larger, the system becomes unstable to spin density wave
formation on top of the superconductivity at the nesting wavevector of the
d-wave nodes.  The case of half-filling has been worked out in detail by
Hsu \cite{hsu} and we only quote the result here. The quasiparticle
dispersion relation Eq. (\ref{dispersion}) becomes modified to
$E_{\bf k} = \sqrt{ ( \epsilon_{\bf k} - \mu )^2 + \Delta_{\bf k}^2
+ \Delta_0^2 }$ , where the energy gap $\Delta_0$ is related to the site
magnetization $< \! \Psi | S_j^z | \Psi \! >/< \! \Psi | \Psi \! > =
\pm m$ by

\begin{equation}
m \simeq \frac{m_0}{1 - \alpha_0 (1 - 4m_0^2)}
\; \; \; \; \; \;
m_0 = \Delta_0/\frac{2}{N} \sum_{\bf k}^N E_{\bf k}
\; \; .
\end{equation}

\noindent
It was observed by Hsu \cite{hsu} that strong projection enhances the
magnetization by roughly a factor of 2 over its unprojected value,
enabling the Heisenberg model value of $m = 0.307$ to be
achieved with a rather modest value of $\Delta_0$, about 1/3 the zone
average of $E_{\bf k}$.  In contrast to the case worked out by him,
however, the gap here has physical meaning and can be detected in a
tunneling or photoemission experiment.  This small quasiparticle gap is a
key characteristic of the gossamer superconductor.

Phase fluctuations have been left out of Eq. (\ref{hamiltonian}) on the
grounds that they are irrelevant to the fermi spectrum, which is
characterized by an energy scale much higher than the superconducting
T$_c$.  However, they are essential for accounting for both the Uemura
plot and strange-metal transport above T$_c$.  The transport of a dilute
gas of bosons formed when the order parameter dephases would not exhibit
any traditional metallic behavior and indeed would tend to ``short out''
conduction by the fermions.  Lattice-mediated crystallization of a
gossamer superconductor would also provide a ready explanation for why the
static stripes \cite{tranquada} are observed at $\delta = 1/8$ is some
cuprates and not others, why the stripe commensuration wavevector is so
peculiar, why stripes can be destroyed by moderate pressure
\cite{pressure}, and why the materials insulate when subjected to
strong magnetic fields \cite{boebinger}.

I wish to express particular thanks to Z.-X. Shen and my other NEDO
collaborators H. Eisaki, A. Fujimori, S. Maekawa, N. Nagaosa, N. P. Ong,
Y. Tokura, and S. Uchida for their collegiality and innumerable helpful
discussions. This work was supported by the Department of Energy under
contract No. DE-AC03-76SF00515.

\end{document}